\documentclass[aps,twocolumn,groupedaddress,showpacs,floatfix,naturefm]{revtex4}

\usepackage{caption}
\usepackage{gensymb}
\usepackage{blindtext}
\usepackage{siunitx}
\usepackage{epsfig}
\usepackage{float}
\usepackage{color}
 \usepackage{ragged2e} 
 \usepackage[utf8]{inputenc}
\usepackage{caption}
\DeclareCaptionJustification{justified}{\justifying}
\usepackage[font=small,labelfont=bf,
   justification=justified,
   format=plain]{caption}
\begin{document}
\title{Observation of $\Delta J$=0 Rotational Excitation in Dense Hydrogens}

\author{Jie Feng$^{1,2}$, Xiao-Di Liu$^{1,*}$, Haian Xu$^{1,3}$, Pu Wang$^{1,2}$, Graeme J. Ackland$^4$ and Eugene Gregoryanz$^{1,3,4,5,*}$}

\begin{abstract}
\noindent
Raman measurements performed on dense H$_2$, D$_2$ and H$_2$+D$_2$  in a wide pressure-temperature range reveal the presence of the 
$\Delta J$=0 rotational excitation.  In the gas/fluid state this excitation has zero Raman shift, but in the solid, 
the crystal field drive s it away from the zero value {\it e.g.}  $\sim$ 75 cm$^{-1}$ at around 50 GPa and 10  K for both isotopes and their mixture. 
In the case of deuterium, the $\Delta J$=0 mode splits upon entering phase II suggesting a very complex molecular 
environment of the broken symmetry phase (BSP).  In the fluid state and phases I and II the frequencies (energies) of 
the $\Delta J$=0 transition for H$_2$ and D$_2$ do not scale either as rotational (by factor of 2) nor vibrational (by $\sqrt{2}$) modes
and appear to be completely isotope independent. This independence on mass marks this transition as unique and a 
fundamentally different type of excitation from the commonly considered harmonic oscillator and quantum rotor. 
\end{abstract}

\affiliation{$^1$Key Laboratory of Materials Physics, Institute of Solid State Physics, HFIPS, Chinese Academy of Sciences, Hefei, China \\
$^2$University of Science and Technology of China, Hefei 230026, China \\
$^3$Shanghai Key Laboratory MFree, Shanghai Advanced Research in Physical Sciences, Shanghai, China \\
$^4$School of Physics and Centre for Science at Extreme Conditions, University of Edinburgh, Edinburgh, UK\\
$^5$Center for High Pressure Science \& Technology Advanced Research,  Shanghai, China \\
$^*$Contact authors: xiaodi@issp.ac.cn, e.gregoryanz@sharps.ac.cn} 
\maketitle

The hydrogen molecule (H$_2$) is a paradigm of a fundamental system in quantum mechanics. Its simplicity,  
combined with the well-defined spectroscopic properties, makes it a textbook example for illustrating key concepts, such as energy 
level quantization and their mixing, selection rules, and the interplay between molecular structure 
and quantum states. It is known that  the Schr{\"{o}dinger equation can be solved exactly only for the hydrogen atom but 
the solutions for a diatomic hydrogen molecule, where the rigid rotor and harmonic oscillator approximations, offer a clear 
framework for understanding discrete energy levels, providing answers about the rotational and vibrational 
excitations \cite{Herzberg_book,Silvera_RMP_1980,Kranendonk_book,Sathyanarayana_book}.
The idealised rotational transitions occur between the levels described by spherical harmonic 
wavefunctions $Y_{Jm_J}$ with angular momentum J, produce a characteristic sequence of states with energy 
spacings that follow a quadratic dependence on the rotational quantum number J and scale between isotopes by 
factor of 2 \cite{Silvera_RMP_1980,Howie_PRL_2012} and can be experimentally observed as such in the gas/fluid state \cite{Pena_JAP_19,Pena_JPCL_2020,Cooke_PRB_2020}. 
The vibrational excitations demonstrate the quantization of molecular bond oscillations, with energy levels that 
(in the simplest approximation) follow a harmonic progression and scale between H$_2$ and D$_2$ by factor of $\sqrt{2}$ \cite{Silvera_RMP_1980,Howie_PRL_2012}. 
The  solid state (density increase) usually leads to the line-shape changes, more complicated appearance of the spectra  
and imposes the restrictions related to the symmetry caused by the crystal field, which could lift the degeneracy of the energy levels. 
The vibrational excitations have energy of around 4200 cm$^{-1}$ (3000 cm$^{-1}$ for D$_2$), which is the highest 
Raman energy of any material and therefore is hardly influenced by the crystal field. The rotational modes'
frequencies are on the order of hundreds of reciprocal centimeters and so are more likely to be affected by the intermolecular 
interactions in the crystalline state. The Raman selection rules for the rotational excitations state that only transitions 
with $\Delta J$=2 and 0 are allowed \cite{Kranendonk_book,Herzberg_book,Silvera_RMP_1980}. $\Delta J$=2 transitions 
are readily observed in H$_2$, D$_2$ and their mixtures 
\cite{Howie_PRL_2012,Pena_JAP_19,Pena_JPCL_2020,Cooke_PRB_2020,Howie_PRL_2014,Liu_PRL_2017,Liu_PNAS_2020,Xu_PRB_2025}.

In the gaseous/fluid state all levels of a given J are degenerate, 
leading to the $\Delta J$=0 Raman shift being zero frequency. But the picture changes in the solid state,  as the 
hexagonal crystal field  lifts the degeneracy of the m$_J$ levels resulting in the $J$=1 splitting into two levels 
($m_J$=0,$\pm 1$, the latter pair being degenerate), $J$=2 into three  ($m_J$=0,$\pm 1$,$\pm 2 $)\cite{Silvera_RMP_1980,Kranendonk_book,Cooke_PRB_2020}.
As well as splitting the roton peaks, lifting the degeneracy  would  lead to a $\Delta J$=0 transition with non zero energy between the m$_J$ states, see 
Fig. S1 in Ref. \cite{Silvera_RMP_1980,SoM}. 
To the best of our knowledge, the $\Delta J$=0 rotational excitation, called "zero roton" here,  of 
molecular H$_2$ was never reported experimentally in the condensed state. We attribute that to the technical difficulties 
of high-pressure Raman spectroscopy {\it e.g.} the zero roton central frequency is close to zero and therefore would 
be masked by the intense  elastic laser line reflected by the diamonds in the standard 180$^{\degree}$ diamond anvil cell Raman experiment. 
The improvements in the light scattering experimental techniques, such as  arrival of the notch filters, which permit 
the observation of Raman excitations close to the laser line $<$ 25 cm$^{-1}$ could allow the investigation of this  
effect. Previously, we have observed unexplained Raman peaks located at 
very low wave numbers (below 100 cm$^{-1}$), see Fig. 1 in Refs. \cite{Liu_PRL_2017,Pena_JPCL_2020}, Fig. 2 in \cite{Pena_JAP_19}
and Fig. 3 in \cite{Liu_PNAS_2020}, suggestive of the $\Delta J$=0 rotational transition. 

\begin{figure*}[t!]
\centering
\includegraphics[width=\linewidth]{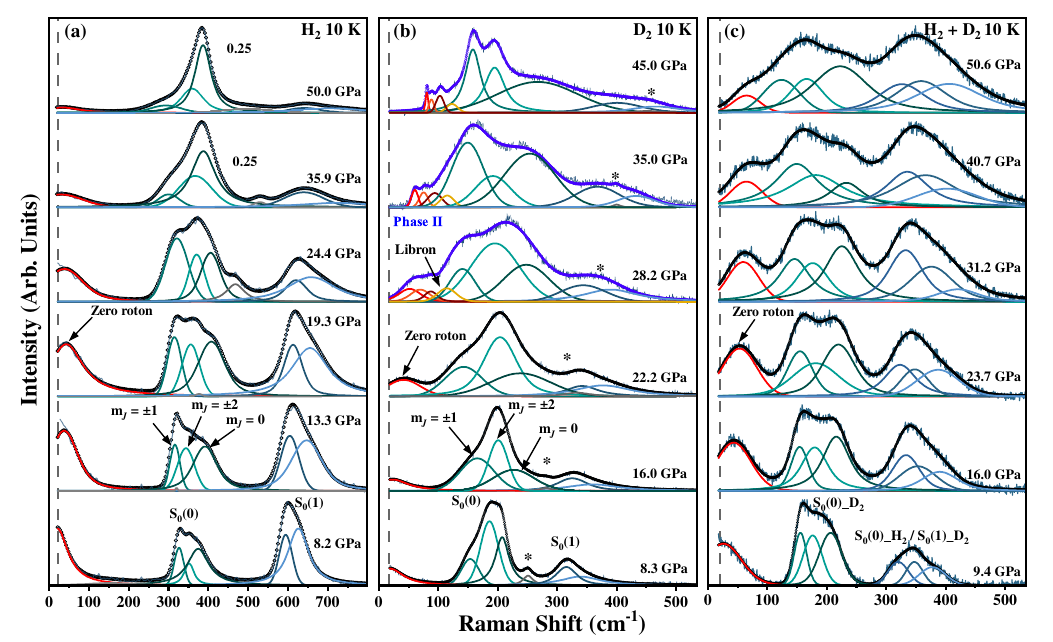}
\caption{Representative Raman spectra of the rotational excitations of H$_2$, D$_2$ and H$_2$+D$_2$ as function 
of pressure at 10 K. (a), (b) and (c) show the evolution of the rotational modes of H$_2$, D$_2$ and H$_2$+D$_2$ respectively. 
The black dotted curves - phase I, blue - phase II for all panels. The fits to the $\Delta J$=0 modes are shown in red in 
all panels and  its split components in D$_2$-II are in orange and dark red.  The new librational mode 
of D$_2$-II is shown in dark yellow.
The S$_0$(0) mode was fitted with 3 peaks corresponding 
to 3 m$_J$ components (see text). Due to the significant overlap and large number of the m$_J$ components
making up the S$_0$(1) mode \cite{SoM}, it was fitted with only 2 or 3 peaks, which have no physical meaning. 
Asterisks ($\ast$) mark the lattice (phonon) mode. 
The vertical dashed line marks the cut-off of the elastic laser line masking the signal.}
\label{fig:1}
\vspace{-15pt}
\end{figure*}
\begin{figure}[h!]
\centering
\includegraphics[width=\linewidth]{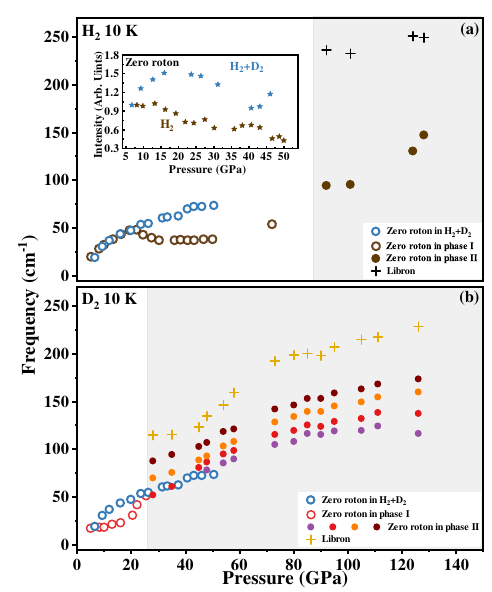}
\caption{Measured frequencies of the $\Delta J$=0 rotational mode as a function of pressure at 10 K.
(a) H$_2$: phase I - the brown empty circles, phase II - the solid circles. Inset: 
the intensity of the zero roton with pressure in H$_2$ and H$_2$+D$_2$.  
Note the frequency decrease in H$_2$ due to the {\it ortho-para} conversion. (b) D$_2$: phase I - the red empty circles, phase II - the solid circles. 
The areas of phase II in D$_2$ are shaded gray. The empty blue circles indicate the frequency of the $\Delta J$=0 mode of H$_2$+D$_2$ in both panels. 
The "+" indicate the librational mode appearing in phase II.}
\label{fig:2}
\vspace{-15pt}
\end{figure}
In this Letter, by combining diamond anvil cell high-pressure technique with Raman spectroscopy, 
we investigate the $\Delta J$=0  transitions in solid H$_2$, D$_2$ up to $\sim$150 GPa and H$_2$+D$_2$ mixture up to 50 GPa in between
10 and 300 K. We observe this transition as the Raman peak with $\omega\approx$0 cm$^{-1}$ in the fluid state, which 
rapidly increases its frequency with pressure at low temperatures. We trace this excitation to 
persist deep inside the stability field of phases II of both isotopes reaching just below 150 cm$^{-1}$ 
at around 125 GPa for H$_2$ and 143 GPa for D$_2$. However, upon entering D$_2$-II at $\sim$25 GPa 
this excitation starts to split, suggesting multiple molecular environments of D$_2$-II.
Meanwhile, in hydrogen its frequency monotonically increases without any splitting} 
through phase I to II transformation at above 70 GPa. At all pressures the   frequency of the 
$\Delta J$=0  transition appears to be isotope independent, suggesting 
a fundamentally different type of excitation from the commonly considered harmonic oscillator and quantum rotor.

For the experimental details, description of the set-up, additional figures, fitting details  and short description 
of the rotational levels of hydrogen/deuterium and {\it ortho-para} conversion physics and H$_2$/D$_2$ mixtures the reader is referred to 
Supplementary on-line materials \cite{SoM}, Refs. \cite{Mao_Calibration_1986,Akahama_JAP_2006,Silvera_RMP_1980,Kranendonk_book,   Liu_PRL_2017,Liu_PNAS_2020,Xu_PRB_2025} and references therein. 
The energy of the  $J$=2 level is 510 K while $J$=1 is 170 K \cite{Silvera_RMP_1980}, 
implying that at 10 K only $J$=1 level could be thermally populated, see Fig. S1. On the other hand, the excitation, coming from the $J$=0 
level and having frequency $\omega$=0, is always present at any {\it P-T} condition but would
always be masked by the laser line due to its close to zero width.
Fig. 1 shows the observed spectra of the rotational modes of H$_2$, D$_2$ and H$_2$+D$_2$ as a function of pressure up to $\sim$50 GPa 
at 10 K, which can only come from the splitting of $J$=1 level (see Fig. S1). The rotational spectra of H$_2$ and D$_2$ above 45 and 
below 150 GPa are shown in Fig. S2. We observe the single zero roton peak with weak isotope 
dependence, appearing to be very symmetric and sharper than $\Delta J$=2 triplet bands 
{\it c.f} the shape of S$_0$(0). For all isotopes its frequency increases rapidly causing it to overlap with the nearest S$_0$(0) by $\sim$30 
GPa for D$_2$, 40 for H$_2$+D$_2$ and above 70 for H$_2$. It is known that the {\it ortho-para} conversion in H$_2$ is strongly 
dependent on pressure, exponentially increasing at above $\sim$10 GPa {\it e.g.} 
at 25 GPa it is 10$^2$ times faster than at 7 GPa \cite{Eggert_PNAS_1999}.  In one of the experiments the H$_2$ sample was kept for 
around 1 hr at $\sim$25 GPa, leading to some {\it ortho-para} conversion \cite{Eggert_PNAS_1999} and de-population 
of the $J$=1 level, as evidenced by the increase in the S$_0(0)$/S$_0(1)$ intensity ratio and shape change as 
well as the decrease of the zero roton intensity, Fig. 1(a). 
With further compression the frequency of the zero roton is monotonically increasing reaching 150 cm$^{-1}$ at 124 GPa, 
deep in the stability field of phase II \cite{Liu_PRL_2017}. 

The D$_2$ {\it ortho-para} spontaneous conversion 
rate is much slower than H$_2$ due to the weaker nuclear spin-rotational coupling. 
Also, due to its larger mass, the D$_2$ rotational modes are more closely spaced in frequency than those of H$_2$, see Figs. 1(a,b).
This leads to the  S$_0$(0) peak starting to overlap the zero roton by around 22-23 GPa, when its central 
frequency reaches above 42 cm$^{-1}$. At around 25 GPa deuterium enters phase II \cite{Liu_PRL_2017} at which moment 
the zero roton splits into 4 rather well defined peaks, Figs. 1(b) and S2. 
We also  note that the pressure where the splitting of different orientations (m$_J$) approaches the 
energy of an excited rotor coincides with the "broken symmetry" phase II: it suggests that the low energy 
of the spherical $J$=0 free rotor ground-state is outweighed by the crystal field.  

\begin{figure}[t]
\includegraphics[width=\linewidth]{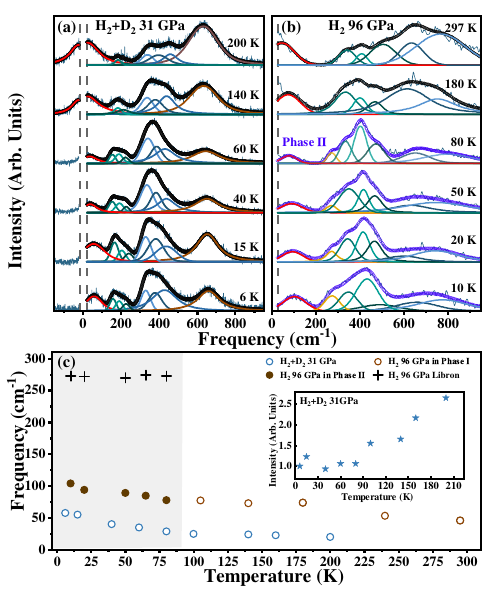} 
\caption{The rotation excitation modes of H$_2$+D$_2$ and H$_2$ as a function of temperature. 
Representative Raman spectra of the H$_2$+D$_2$ mixture at 31 GPa (a) and H$_2$ at 96 GPa (b) over a wide temperature range. The black dotted curves - phase I, 
blue - phase II for (a) and (b) panels. (c) The frequency of the zero roton mode as function of temperature at 31 and 96 GPa.  D$_2$: phase I - the 
brown empty circles, phase II - the solid circles. The areas of phase II in D$_2$ are shaded gray. H$_2$+D$_2$: the empty blue circles. 
Inset: The intensity of zero roton in H$_2$+D$_2$ at 31GPa.  }
\label{fig:3}
\vspace{-15pt}
\end{figure}
Fig. 1(c) shows the results of an experiment on H$_2$+D$_2$ 50:50 mixture, which were loaded by condensing 
both gases at 6-10 K, see Supplementary Materials \cite{SoM}. Loading H$_2$ and D$_2$ at 300 K always leads to the 
collision-based formation of HD \cite{Howie_PRL_2014,Liu_PNAS_2020,Xu_PRB_2025}, but at 10 K the thermal motion 
is negligible, thus preventing the reaction leaving the overall spectra relatively simple.
The mixture represents quite an interesting system, that consists of 2 chemically identical particles having very 
different masses. It  appears that the presence of the  both isotopes not only slows down (at least on the time scale of the 
experiment) the {\it ortho-para} conversion in hydrogen but also prevents the formation of phase II of D$_2$ 
at $\sim$24-25 GPa \cite{Liu_PRL_2017}.  We note that when 3 species (including HD) are present in the sample, 
phase II is delayed to above 60 GPa \cite{Liu_PNAS_2020}. Therefore, despite the apparent complication caused by the second 
isotope, the mixture allows us to circumvent both conversion and phase II formation and
observe the undisturbed zero roton up to 50 GPa. 
In the H$_2$+D$_2$ system, the zero roton peak frequency increases monotonically with pressure, 
reaching 75\,cm$^{-1}$ at 50 GPa. The zero roton peak is resolved up to 35 GPa and  can be easily traced as the lower frequency shoulder 
to at least 50 GPa, Fig. 1(c).  Presumably, this peak has contributions from both D$_2$ and H$_2$, but it is very symmetric, 
suggesting that the zero roton has the same frequency in both isotopes. The same frequency of the zero 
roton for the pure isotopes and their mixtures imply the mass independence for this excitation. We also note that the 
relative intensity of the zero roton appears to be the weakest in the case of D$_2$, probably due to the relative population of the rotational levels. 

Below 25 GPa the S$_0$(0) peak of pure H$_2$ looks similar to that one of D$_2$ in mixture, while S$_0$(0) peak of 
pure D$_2$ is similar to that of hydrogen but at much higher pressures.  This supports the idea that the {\it ortho-para} conversion 
is absent in mixtures, providing an example of the peak shape not influenced by the conversion.
 
Fig. 2 presents the frequencies of the zero rotons for two isotopes and their mixture. 
The evolution of the rotational hydrogen modes frequency and intensity represents a special case due to the  {\it ortho-para} 
time dependent conversion \cite{Eggert_PNAS_1999}, Fig. 2(a). It causes the depletion of the $J$=1 level, 
decreasing the intensity of the zero roton within first 50 GPa, {\it c.f.} to the intensity of the zero roton of H$_2$+D$_2$,
inset of Fig. 2(a). This  leads to unique configuration, where the $\Delta J$=0 at $J$=0 process is dominant. Since this 
process has zero frequency it drives also the frequency of the zero roton down. The frequency of  
deuterium's zero roton, is lower than that of H$_2$ below 20 GPa but reaches exactly  the same values at higher compressions, 
Fig. 2(b). When phase II is reached, the zero roton splits (see above) which could be traced up 
to 143 GPa at low temperature, as shown in Figs. S2 and S3. The highest frequency band which originates from the $S_0(0)$, Figs. 1 and S2,
is very close in frequency to the zero roton but in fact is the librational mode.

Neither the structure of H$_2$-II or D$_2$-II is  known \cite{Goncharenko_Nat_05} but it is assumed that 
D$_2$-II would have lower symmetry  \cite{Geneste_PRL_12}. We speculate that when D$_2$-I transforms from  hcp  with one molecular
cite to phase II, it acquires additional  
distinct molecular environments \cite{Goncharenko_Nat_05,pickard2009structures,van2020quadrupole,zong2020understanding}.
In this case, each new peak would come from the $J$=1 level of molecules belonging to a different Wyckoff site.
The structure of D$_2$-II due to the orientational ordering is 
expected to be more complex than that of H$_2$ \cite{Liu_PRL_2017} {\it e.g.} having more 
unique molecular sites. It was suggested that quantum effects favor molecular rotation
thus decreasing the symmetry of D$_2$-II compared to that of H$_2$-II \cite{Geneste_PRL_12}. This, as well as 
previous optical studies \cite{Goncharov_PRB_96,Liu_PRL_2017}, seem to support this idea.
It is known that the rotational modes, lattice phonon and fundamental vibrational mode $\nu_1$ 
could interact with each other and produce the combinational bands and/or overtones \cite{Hanfland_PRL_92,Eggert_PRL_93,Mao_RMP_94}. 
The multitude of peaks previously reported in the D$_2$-II phase (see Figs. 1 in Refs. \cite{Goncharov_PRB_96, Liu_PRL_2017}) 
can be attributed to a combination of the $Q_1$(1)-$\nu_1$ mode and the zero rotons S$_0$(1)$_{m_1}$
reported here (see Figs. S2, S3 in \cite{SoM}). 

The zero roton frequency in the mixture, Fig. 2(a,b) reproduces that
one in H$_2$ up to $\sim$20 GPa and is very close to the one of D$_2$ at higher pressures, thus appearing  as mass independent and not scalable 
as other H/D excitations. We argue that the frequency observed in the mixture represents the 
unperturbed frequency of the zero roton both for H$_2$ and D$_2$ otherwise modified by some other 
processes such as {\it ortho-para} conversion of phase transition.

The quantitative analysis of the frequencies poses an interesting question about the splitting of 
the rotational J levels. At 10 K the S$_0(0)$ and S$_1(0)$ rotational bands are present, but S$_0(0)$ is the 
only rotational excitation in which the splitting of the rotational level ($J$=2) can be deduced (at moderate pressures) 
from the fitting to its 3 individual components, see Fig. 1 and 
Ref. \cite{Pena_JPCL_2020}. We have fitted the observed S$_0$(0) mode with 3 peaks, which correspond to 
$m_J$=$\pm$1,2,0, see Fig. S1 and also Fig. 4 in Ref. \cite{Cooke_PRB_2020}.  Curiously, the frequency 
difference between $|\Delta m_J|$=$|0|-|2|$ states very closely follows the values of the measured 
zero-roton frequency, see Fig. S4 in \cite{SoM}.  However, in this scenario
the zero roton should appear as a triplet roughly mimicking the shape of the S$_0(0)$ while we 
observe one very symmetric peak {\it c.f.} the shape of the zero roton and S$_0(0)$ at lower 
pressures in Fig. 1. A more plausible explanation is that at moderate pressures the splitting 
of the different J levels is roughly similar {\it e.g.} $|\Delta m_J|$=$|0|-|2|$ of $J$=2 $\sim$
$|\Delta m_J|$=$|0|-|1|$ of $J$=1 but would start to deviate as pressure is increased. 

Fig. 3(a,b) demonstrates the Raman spectra as a function of temperature at 31 GPa for the H$_2$+D$_2$ 
and H$_2$ at 96 GPa,  while Fig. 3(c) presents their frequencies {\it versus} temperature. 
As temperature goes up, the higher-lying rotational states ($J>1$)
start to get thermally occupied  and therefore contribute additional zero rotons. At higher temperatures 
the contributions from the anti-Stokes excitation start to appear as, Fig. 3(a) see also Fig. S5. Due to the activation of 
the levels such as $J$=2, which would have 3 transitions contributing to the zero roton, and $J$=3 (14 transitions) the resulting peak would 
contains multiple Raman modes and would expect not only to increase in intensity but also to broaden significantly, 
thus resulting in a single broad peak observed in the experiment. As temperatures increase, the "quantum rotor" state moves
towards the "classical rotor" regime, which intensifies the molecular thermal motion with the local molecular environments becoming more homogeneous, 
thereby reducing the J levels splitting thus leading to the softening of the zero roton frequency, Fig. 3(c).

The $\Delta J$=0 rotational transition is the last previously-unobserved excitation in molecular hydrogen which follows 
from the selection rules imposed by quantum mechanics. Its physical properties appear to be unique:
the independence of frequency on mass marks the zero roton as a fundamentally 
different type of excitation from the commonly considered harmonic oscillator and quantum rotor and from that 
point of view  the "zero roton" is perhaps a misnomer.  Raman spectroscopy demonstrates that the observed frequency
is essentially mass independent, as one would expect from a transition between levels with equal kinetic energy.  Although 
the quantum mechanics of the $\Delta J$=0 excitation is well described in terms of perturbed rotational transitions, 
it is more intuitive to think of it as a transition between two in-equivalent 
orientations of the hydrogen molecule. The observed properties of the $\Delta J$=0 excitation, the interplay 
between its intensity, frequency, its behavior during the {\it ortho-para} conversion and the absence of 
the latter in the mixed species, all of it represents a fascinating example of the fundamental quantum mechanics, 
which hydrogen has to offer.

This work was supported by research grants from the National Science Foundation of China (Grant Nos. 12522401, 12204484), 
Innovation Program for Quantum Science and Technology (No. 2024ZD0302100), the Youth Innovation Promotion Association 
of CAS (Grant No. 2021446) and the HFIPS Director’s Fund of Chinese Academy of Sciences (Nos. BJPY2023B02).

\end{document}